# Resource Allocation in IRS-aided Optical Wireless Communication Systems

Ahrar N. Hamad[1], Ahmad Adnan Qidan[2], Taisir E.H. El-Gorashi[2]
and Jaafar M. H. Elmirghani[2]

[1]School of Electronics and Electrical Engineering, University of Leeds, Leeds, United Kingdom,
[2]Department of Engineering, Kings' College London, London, United Kingdom
sc18ah@leeds.ac.uk, ahmad.qidan@kcl.ac.uk, taisir.elgorashi@kcl.ac.uk, jaafar.elmirghani@kcl.ac.uk

**Abstract**
One of the main challenges facing optical wireless communication (OWC) systems is service disconnection in high blockage probability scenarios where users might lose the line of sight (LoS) connection with their corresponding access points (APs). In this work, we study the deployment of passive reflecting surfaces referred to as Intelligent Reflecting Surfaces (IRSs) in indoor visible light communication (VLC) to boost users signal to noise ratio (SNR) and ensure service continuity. We formulate an optimization problem to allocate APs and the mirrors of IRSs to users such that the sum rate is increased. The results show a 35% increase in the sum rate of the IRS-aided OWC system compared to the sum rate achieved by only considering the LoS channel components. The results also shows that the deployment of IRSs improves the sum rate under LoS blockage.

## 1. INTRODUCTION

In recent years, the need for high-speed wireless connectivity has grown significantly leading researchers from both academia and industry to investigate new technologies capable of supporting the escalating traffic of data intensive internet-based applications [1], [2]. In this context, optical wireless communication (OWC) has emerged as a promising 6G technology complementing radio frequency (RF) wireless systems to relieve the spectrum shortage [3]–[6]. The large, unregulated optical spectrum bandwidth has the potential to support data rates beyond 20 Gbit/s per user. Other advantages of OWC include high energy efficiency, enriched security and low cost [7]–[12]. However, OWC faces a number of challenges including inter-symbol interference (ISI) caused by multipath propagation and power constraints imposed by eye and skin health and safety. An angle diversity receiver (ADR) composed of multiple photodiodes, each with narrow field of view (FoV) [7], [8], [13], can be used to overcome Inter Symbol Interference (ISI) while maintaining users' connectivity. OWC systems can also suffer from severe performance degradation due to line of sight (LoS) blockage by objects in the environments. Moreover, in visible light communication (VLC) [14], where LED-based optical APs are used for illumination and communication, the confined coverage area of APs results in the need for a large number of APs to ensure full coverage.

Recently, Intelligent Reflecting Surfaces (IRSs) have been proposed to expand the connectivity of RF networks, while using a limited number of base stations, by focusing the base station signals towards users to reduce ISI and overcome LoS blockage. An IRS is composed of passive reflecting elements made from metasurfaces or mirrors. Typically, these reflecting elements are arranged in a planar surface to independently reflect the incident signals at different angles by different amplitude and/or phase [15].The deployment of IRS in conventional RF networks was shown to enhance performance in terms of spectral and energy efficiency [16].

In the context of indoor OWC, IRSs can be deployed as a key solution to ease the impact of LoS blockage, broaden the coverage and enhance the achievable user rate. In [17], the use of multiple reconfigurable intelligent surfaces in OWC was shown to improve the system performance by reducing the outage probability. In [18], the performance of meta surface and mirror array-based reflectors is studied in an OWC system and the results show that the power received by users is determined by the number of reflecting elements and their orientation and position. In addition, a novel physical layer security technique for an IRS-aided indoor OWC system is proposed in [19]. The orientation of the mirrors is optimized such that the IRS-based optical channel of the legitimate user is enhanced, while ensuring that eavesdroppers experience a weak IRS-based optical channel. Moreover, the optimization of IRS reflection coefficient using greedy algorithm to maximize the sum rate was considered in [20]. It is also shown that IRS based beam steering can improve signal reception in VLC systems by steering the incident light beam using meta lens and crystal liquid [21].

In this paper, we investigate the improvement of sum rate in mirror-based IRS-aided OWC systems. We optimize the allocation of APs and mirrors to users such that the sum rate of users is maximized. We solve the optimization problem using exhaustive search and compare the sum rate in the IRS-aided OWC system to the sum rate achievable by the LoS channel components only and LoS and diffuse non-line-of-sight (NLoS) channel components under varying transmitted optical power and LoS blockage ratios. The rest of this paper is organized as follows: In Section 2, the system model is presented. The simulation results are given and discussed in Section 3. Finally, conclusions are presented in Section 4.

## 2. SYSTEM MODEL

A downlink VLC system is considered in an indoor environment with $L$ multi-LED-based optical APs deployed on the ceiling of the room to provide illumination and communication for $K$ users distributed randomly on the communication plane, as shown in Fig. 1 (a). Each AP is composed of multiple LED transmitters to ensure an expanded coverage area under eye safety power constraints. Each user is equipped with an ADR set to ensure that the user can be served by all the APs in the room. The direction of the ADR photodiodes is determined by their azimuth angle ($AZ$) and the elevation angle ($EL$). The walls, ceiling, and floor are modelled as Lambertian reflectors. More information on the calculation of the channel responses can be found in [22]–[28]. The reflected signals from the walls, ceiling and floor of the room cannot be controlled to maximize the gain of users. To enhance the SNR, mirror arrays are mounted on the walls to act as IRSs specular reflecting the APs signal to the APs assigned users. Each array of mirrors is composed of $width_m \times hight_m$ identical, passive, smooth reflecting rotational $m$ mirrors. Each mirror is fixed to a randomly selected rotation determined by two independent angles; the roll angle around the x-axis and the yaw angle around the z-axis, as shown in Fig. 1 (b). Note that, given the large number of mirrors and the limited area of the indoor environment, random rotations of mirrors will most probably result in full coverage of the room, i.e., each user will most probably find a mirror that enhances its received signal.

Typically, OWC is intensity modulation/direct detection (IM/DD). Hence, on-off Key (OOK) modulation is considered to avoid complexity. All the APs are connected to a central unit (CU) that collects information on the network status in terms of resource availability and users demands. The CU uses this information to optimize the allocation of APs and mirrors to users such that the users sum rate is maximized. The optical channel is composed of a LoS component, which is a direct link from the AP to the user, and NLoS components made of diffuse reflections by objects in the environment. In this work, without loss of generality, we model up to the second reflections, ignoring higher reflections components of the NLoS channel. In addition to the LoS and diffuse NLoS components, the users in an IRS-aided -OWC system receive mirror-reflected components. Note that in this work we only consider first reflections by the mirrors, i.e., a signal from an AP reflected by a mirror to a user. The signal received by user $k$ from optical AP $l$ can be written as

$$P_k = P_t [ O_{k,l} h_k^{LOS} + h_k^{NLOS} + h_k^{IRS} ] + n_k, \qquad (1)$$

where $P_t$ is the transmitted power, $h_k^{LOS}$ is the LoS channel impulse response, $h_k^{NLOS}$ is the impulse response of the NLoS channel components reflected by the walls, ceiling and floor, $h_k^{IRS}$ is the IRS channel impulse response of the reflected components by the mirrors within the array, $n_k$ is the real-valued Additive White Gaussian Noise (AWGN) of user $k$ with zero mean and variance and $O_{k,l}$ is a binary variable $O_{k,l} = 1$ if there is no obstacle between user $k$ and AP $l$, otherwise $O_{k,l} = 0$.

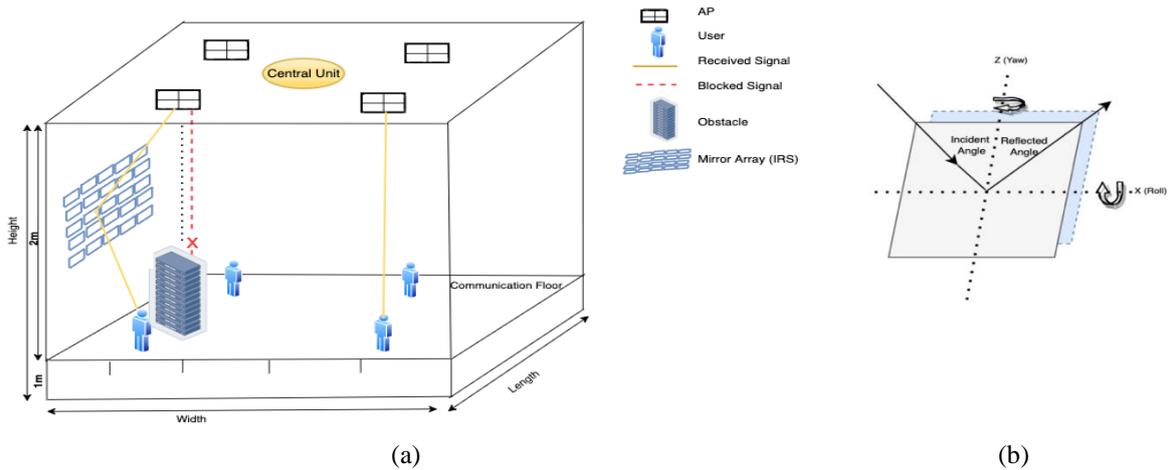

*Figure 1. (a) IRS-aided downlink VLC system model, (b) Rotational Mirror.*

## 3. SIMULATION RESULTS

To demonstrate the improvement in OWC systems sum rate obtained by optimizing resource allocation in an IRS-aided OWC system, we consider the system model with room dimensions of $5m \times 5m \times 3m$ with four LED-based APs ($L = 4$) deployed on the ceiling and two mirror arrays mounted on opposite walls. Each mirror array contains $5 \times 5$ reflective elements, each with an effective area of $25\ cm \times 15\ cm$. As mentioned earlier, each mirror is fixed to a randomly selected rotation. On the receiving plane, four active users ($K = 4$) are randomly distributed. Other simulation parameters are listed in Table 1.

An optimization problem is formulated to allocate resources among APs and mirrors in each mirror array to users, with the objective of maximizing the aggregate sum rate of all users. It is worth mentioning that the resources considered in this work are assigned on a fractional time basis, i.e., resources are devoted to a user for the time required to send its data. In this context, a utility-based objective function in a logarithmic form is defined to maximize the sum rate and to preserve proportional fairness among users. The optimization problem is solved through exhaustive search.

*Table 1: SYSTEM PARAMETERS* [29]

| Parameter | Value | | | |
|---|---|---|---|---|
| **Room Configurations** | | | | |
| Length x Width x Height | 5 x 5 x 3 m$^3$ | | | |
| Reflectivity of walls, floor, ceiling | 0.8,   0.3,   0.8 | | | |
| Area of diffuse reflecting element | 1st | | 2nd | |
| | 5 cm x 5 cm | | 20 cm x 20 cm | |
| **LED Transmitter** | | | | |
| Quantity | 4 | | | |
| Location | (1.5,1.5,3), (1.5,3.5,3), (3.5,1.5,3), (3.5,3.5,3) | | | |
| Transmitted power | 2 W | | | |
| Half power semi-angle | 60° | | | |
| **MA-IRS** | | | | |
| Number of Mirror array | 2 | | | |
| Reflectivity of Mirror | 0.95 | | | |
| **ADR Receiver** | | | | |
| Quantity | 4 | | | |
| Responsivity | 0.4 A/W | | | |
| Physical area in a PD | 20 $mm^2$ | | | |
| Bandwidth | 20 MHz | | | |
| ADR Photodetector Branches | 1 | 2 | 3 | 4 |
| Azimuth angels | 0° | 90° | 180° | 270° |
| Elevation angels | 60° | 60° | 60° | 60° |
| Field of View | 25° | 25° | 25° | 25° |

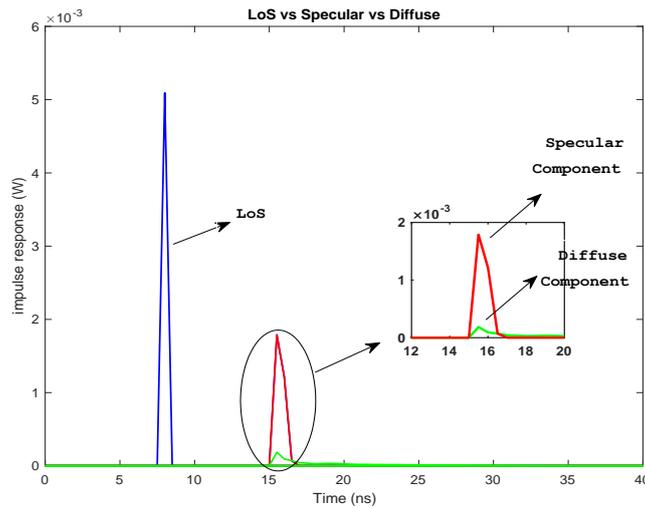

*Figure 2. Impulse channel responses for a user located in the centre of the room.*

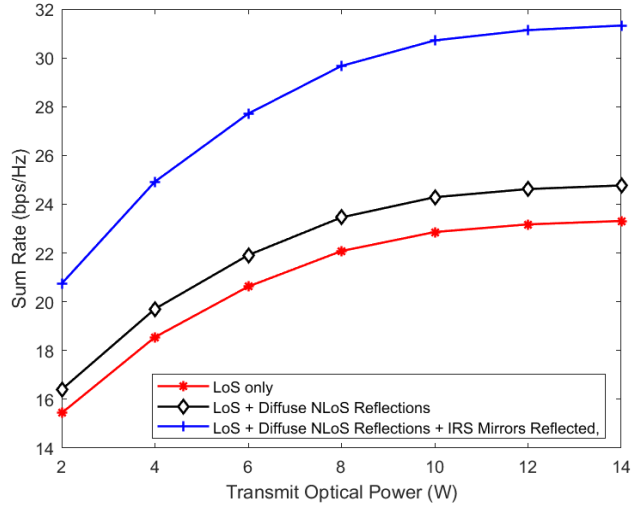

*Figure 3. Sum rate versus transmit optical power.*

In Fig. 2, the impulse channel responses of LoS, diffuse NLoS and IRS-NLoS specular reflected components are depicted for a user located in the center of the room to illustrate their power contributions. It can be seen that the power contribution of the LoS component is significant compared to the diffuse components, which means that if a user loses the LoS link with its corresponding AP, it experiences very low SNR, and therefore the deployment of IRS is essential in OWC. Moreover, the reflected signal from the mirror has higher power compared to the power contributions of the reflected signals from the walls, ceiling and floor due to the ability of IRS to focus the reflected signals towards users.

Figure. 3 compares the IRS-aided OWC system sum rate to the sum rate achievable by the LoS channel components only and LoS and diffuse non-line-of-sight (NLoS) channels. The figure shows that the sum rate increases as the transmitted optical power increases. However, the transmitted power is subject to illumination and eye safety regulations. Deploying the IRS improved the sum rate by 35% and 29% compared to LoS only and LoS and diffused NLoS, respectively.

Figure. 4 shows the sum rate against the LoS blockage ratio. As expected, the achievable rate decreases as the LoS blockage ratio increases. However, the use of IRS relieves the blockage problem in OWC. The use of two 5x5 mirror arrays achieves a sum rate up to 4.8 bps/Hz at a blockage ratio of 1 while the sum rate is limited to 1.8 bps/Hz considering the LoS and diffuse NLoS components. The results also show improvement in the sum rate (2.4 bps/Hz at a blockage ratio of 1) by using a single 5x5 mirror array compared to LoS only.

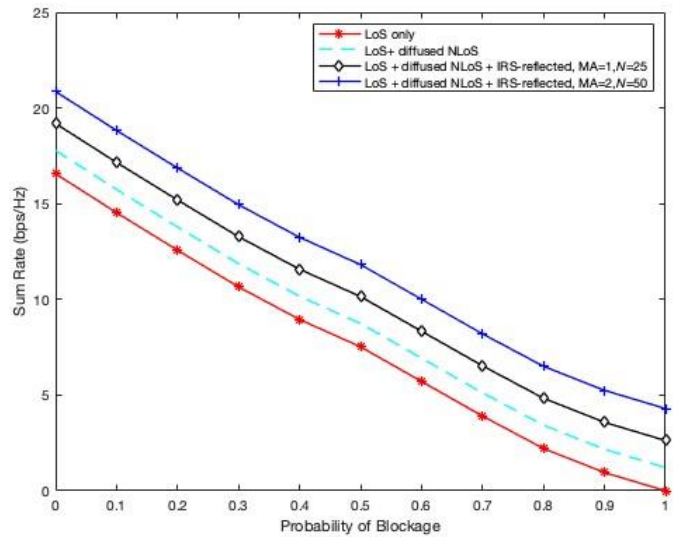

*Figure 4. Sum rate versus blockage ratio.*

**4. CONCLUSIONS**

This paper investigated the performance of IRS-aided indoor OWC system composed of multiple LED-based APs serving multiple users and arrays of rotational mirrors acting as the IRSs. An optimization problem was formulated to maximize the sum rate by optimum allocation of APs and IRS mirrors. The allocation of APs and mirrors is solved through exhaustive search. To relax complexity, APs are first allocated to users followed by mirror allocation to users. The results show that deploying two 5x5 mirror arrays increases the sum rate by 35% and 30% compared to the sum rate achieved by the LoS channel component only and the LoS and diffuse NLoS channel components, respectively. The results also shows that the deployment of mirror-based IRSs improves the sum rate under LoS blockage.

## ACKNOWLEDGEMENTS

This work has been supported in part by the Engineering and Physical Sciences Research Council (EPSRC), in part by the INTERNET project under Grant EP/H040536/1, and in part by the STAR project under Grant EP/K016873/1 and in part by the TOWS project under Grant EP/S016570/1. All data are provided in full in the results section of this paper.